# Ultrafast Photomodulation Spectroscopy: a device-level tool for characterizing the flow of light in integrated photonic circuits

Roman Bruck[1*], Ben Mills[2], David J. Thomson[2], Frederic Y. Gardes[2], Youfang Hu[2], Graham T. Reed[2] and Otto L. Muskens[1]

[1]*Physics and Astronomy, Faculty of Physical Sciences and Engineering, University of Southampton, Southampton SO17 1BJ, UK*
[2]*Optoelectronics Research Centre, University of Southampton, Southampton SO17 1BJ,*
email: r.bruck@soton.ac.uk

**Abstract:** Advances in silicon photonics have resulted in rapidly increasing complexity of integrated circuits. New methods are desirable that allow direct characterization of individual optical components in-situ, without the need for additional fabrication steps or test structures. Here, we present a new device-level method for characterization of photonic chips based on a highly localized modulation in the device using pulsed laser excitation. Optical pumping perturbs the refractive index of silicon, providing a spatially and temporally localized modulation in the transmitted light enabling time- and frequency-resolved imaging. We demonstrate the versatility of this all-optical modulation technique in imaging and in quantitative characterization of a variety of properties of silicon photonic devices, ranging from group indices in waveguides, quality factors of a ring resonator to the mode structure of a multimode interference device. Ultrafast photomodulation spectroscopy provides important information on devices of complex design, and is easily applicable for testing on the device-level.

Integrated silicon-based photonics has developed into a mature technology platform with a multitude of applications [1-4], including telecommunications, healthcare diagnostics and optical sensors. As technology progresses, device designs become increasingly complex and integrate more functions onto a single device. Characterization of fabricated devices is an important step in the design cycle as it highlights differences between the intended design and the fabricated device, thus allowing the optimization of fabrication steps as well as of the entire design process. The toolbox is well filled when it comes to geometrical characterization of waveguides and devices thereof. Established technologies, such as scanning electron microscopy (SEM), atomic force microscopy (AFM), ellipsometry and many others, are able to precisely measure device footprints, waveguide cross sections, surface and sidewall roughness, film thicknesses, and other geometric parameters.

Direct access to properties of light propagation in waveguide devices proves more challenging. Researchers have been inventive and have proposed numerous methods for analysis of integrated optic elements. Among them are reflectometry methods [5-8] to find defects in waveguides, far-field scattering microscopy to measure propagation losses [9-12], as well as interrogation of structures with electron beams (cathodoluminescence) [13], near-field optical probes [13-15], and AFM tips [16-20]. Near-field scanning optical microscopy (NSOM) is a powerful technique that gives direct access to light propagation, including phase information, and features a high spatial resolution [14,15,21,22]. Drawbacks of scanning probe microscopy are the small field of view, slow scanning speeds, and limited reproducibility and durability of the tips. Moreover, near-field techniques require direct access to the waveguide surface in order to interact with evanescent field components, limiting the analysis of devices covered with a top cladding for protection and stability.

Here, we present a new approach for characterization of silicon on insulator (SOI) waveguide elements at the device-level based on non-contact perturbation of the structure using a tightly focused femtosecond laser pulses. The principle of Ultrafast Photomodulation Spectroscopy (UPMS) is shown in Fig. 1. Ultrashort UV laser pulses are focused from the top onto a small area of the device under test (DUT). The pulsed laser excitation produces a highly localized and time-dependent perturbation in the refractive index of silicon, which acts as a local gate for monitoring the flow of probe light through the structure. Hence the timing and the position of this gate is known, it can be used to reconstruct a space and time dependent photomodulation map of the structure by monitoring the transmission through the device. The technique can be used for devices covered with any top cladding that does not absorb the pump pulses, which includes popular choices such as silicon dioxide and most polymer claddings. Ultrafast response of waveguides was studied in the context of adiabatic frequency conversion and ultrafast switching [23-30]. Collinear infrared pump-probe arrangements have been used to characterize pulse propagation and waveguide nonlinearities [31]. While our setup can also by applied for ultrafast control of SOI elements, we focus here on the imaging and quantitative characterization aspect, which is ideally suited for device performance characterization.

**Results and Discussion**

The UPMS technique provides hyperspectral information which can be independently analysed as function of four different parameters, namely two coordinates of the pump spot, wavelength and delay time. Two dimensional scans of the pump spot on the DUT allow for spatial mapping of the mode distribution similar to AFM-based modulation techniques [18]. The spatial resolution of the UPMS is sub-micrometer, and thus comparable to telecom wavelengths inside high-index waveguides. It is limited by the spot size of the pump pulses, as spatial broadening of the exited spot can be neglected for small delay time (<100ps). Hyperspectral information is obtained by spectrally resolving the transmission spectrum over the bandwidth of the femtosecond probe laser. The probe features a Gaussian spectrum of about 44 nm full-width at half maximum. Because of the high sensitivity of the equipment used, a wider wavelength window into the wings of the pulse spectrum can be analysed. Typically, wavelength windows of 60 nm to 100 nm width can be achieved. Finally, the delay time, i.e. the timing between the pump and probe pulses, is used to extract information on the time-response of the structures with ultrashort resolution (150 fs).

**Characterization of refractive index modulation.** As a first step, we quantified the perturbation in the refractive index $\Delta n_{Si}$ caused by pulsed laser excitation employing an asymmetric Mach-Zehnder interferometer (MZI) as the DUT as shown in Fig. 2a. Pulsed laser excitation in one arm of the MZI resulted in a change in the spectral position and the contrast of the interference fringes of the transmission function (Fig. 2b). Its analysis gives simultaneous access to the changes in the real and imaginary part of the refractive index, respectively through the wavelengths shift $\Delta\lambda$ and the fringe contrast (see Methods).

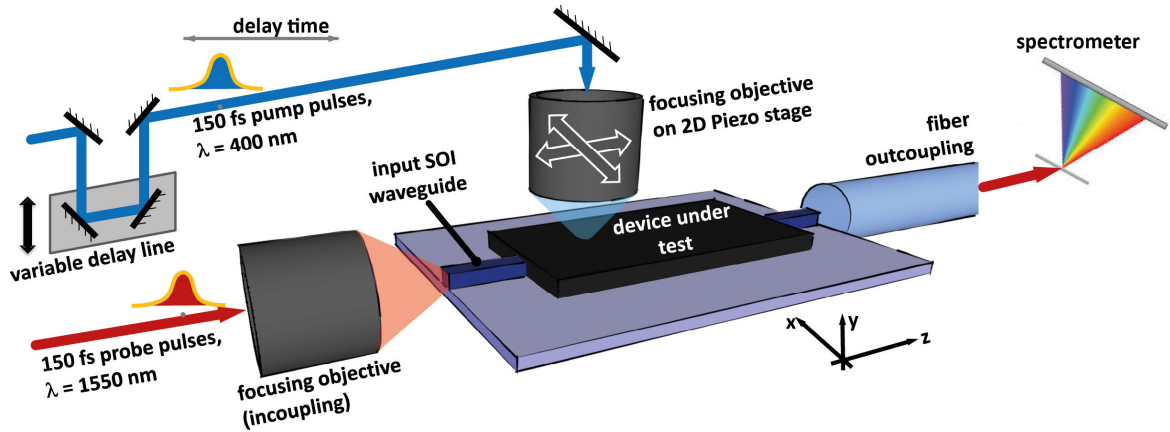

Fig. 1. Schematic of the UPMS setup. Infrared probe pulses are coupled into a SOI waveguide, travel through the device-under-test and are outcoupled to a singlemode fiber. The transmission is spectrally resolved in a spectrometer. Pump pulses are incident on the device region from the top. The incident position of the pump pulses can be scanned over the device and the delay time can be adjusted.

Figure 2c and d display the measured results for $\Delta n_{Si}$ averaged over a wavelength window of 40 nm width around a center wavelength of 1550 nm. Remarkably, the real part of the refractive index decreases by more than 0.4 without any visible damage to the waveguide. As a result, phase shifts greater than $2\pi$ are obtained over a modulated zone of only a few optical wavelengths in length. Some saturation is observed for the change in the real part at high laser powers. The imaginary part of the refractive index, responsible for the induced absorption, increases much less and shows a linear behaviour as function of the pump fluence. For large delay times (not shown in Fig 2d), the real part of $\Delta n_{Si}$ decays exponentially with a time constant of about 330 ps for the rib waveguides under investigation. The imaginary part shows a distinct peak which decays considerably faster.

The refractive index modulation in our experiment is attributed to free carrier contributions. Free carrier nonlinearity is caused by formation of electron-hole pairs and provides a reduction of the refractive index and an increase in absorption [32]. The pump fluence of 100 pJ/$\mu m^2$ has been shown before to result in electron-hole pair densities exceeding $10^{20}$ cm$^{-3}$ [33], for which the free carrier nonlinearity of Si [32] predicts a refractive index modulation of $\Delta n_{Si}$ =-0.173+10$^{-4}$i. Our values for the real part of $\Delta n_{Si}$ are in agreement with these high carrier densities. For long delay times (>20ps), we expect the imaginary part to follow this prediction as well and decay with the same time constant as the real part. Measurements of imaginary parts in the range of 10$^{-4}$, however, are beyond the sensitivity limits of our experiment. For short time delays, higher values than predicted are found for the imaginary part of $\Delta n_{Si}$. Possible additional contributions that could increase transient losses include changes in Drude damping or effective mass [33,34] at high pump fluences and UV wavelengths. Additionally, the presence of surface states in the etched SOI waveguides may result in additional defect-related contributions to absorption and differences in carrier dynamics. Our results are in qualitative agreement with previous studies on switching of Si photonic crystal membranes, which also identified free carrier effects as the dominant contribution [25,35]. Those studies were done at lower pump intensities, resulting in smaller values of $\Delta n$. The low repetition rate in our studies resulted in a negligible thermo-optic background, therefore we have good access to events occurring at negative delay times, i.e. when probe pulses arrive before the pump pulses. This feature is important for time-resolved investigation of resonator elements as will be shown below. The dynamics of the recovery process is governed by relaxation and transverse diffusion of excited carriers. Transverse diffusion causes spatial broadening of the area of effect. We quantified in our experiments that the effect length increases by 500 nm over 100 ps, corresponding to a lateral diffusion constant of around 1 cm$^2$/s. For measurements on timescales of maximum tens of picoseconds, this expansion of the excited spot can therefore be neglected and the resolution is set by the optical excitation area.

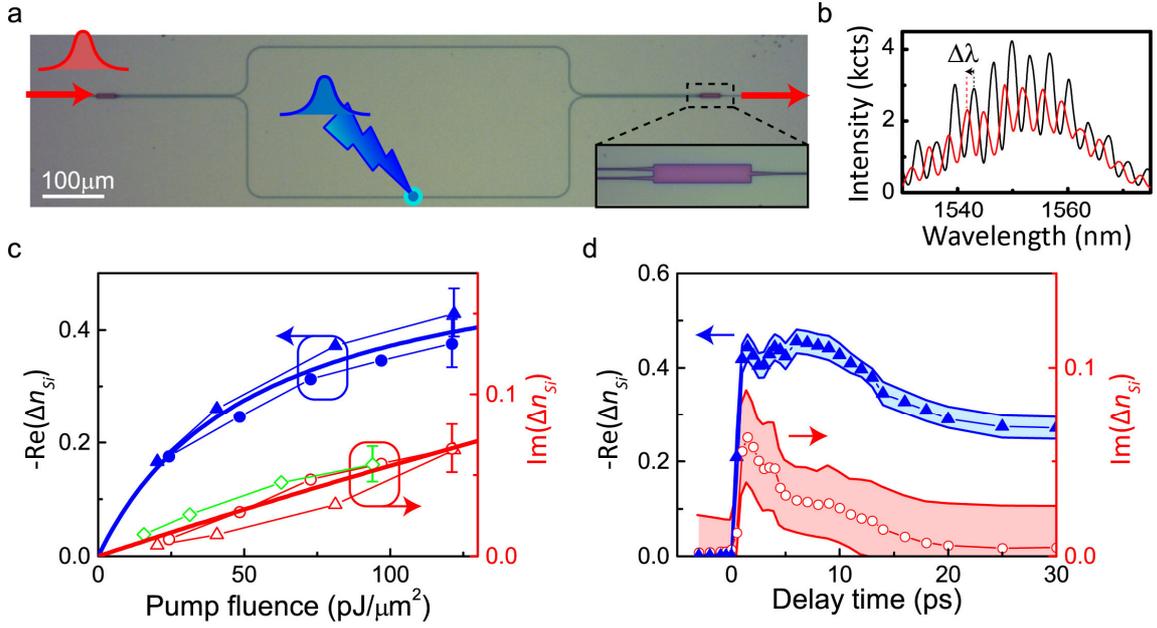

Fig. 2. Quantification of the complex refractive index change in silicon. a) Asymmetric Mach-Zehnder interferometer (MZI) used to determine the refractive index change induced in silicon by the UV pump pulses. b) Transmission spectra (unit kilo-counts) of the unpumped (black) and pumped (red) MZI, showing shift Δλ of interference fringes and reduced fringe contrast due to absorption. c) Measured change of the real (blue) and imaginary (red) part of the refractive index of silicon as function of the pump fluence, for two effect lengths of 1.5 µm (dots) and 3.5 µm (triangles) and a delay time of 5 ps. Green diamonds are direct measurements of absorption through intensity modulation of a straight waveguide. Lines are guides to the eye using linear saturation equation (see Methods). d) Change in the real (blue) and imaginary (red) part of the refractive index as function of delay time (pump fluence 130 pJ/µm², 1.3 µm effect length).

**Group index dispersion of a waveguide.** UPMS allows direct measurements of the group index and its dispersion on any SOI waveguide by precise time of flight measurements using excitation of the waveguide as an ultrafast nonlinear gate (Fig. 3a). So far, group index measurements have been performed on specially designed test elements, such as asymmetric MZIs [36], where the group index has been deducted from transmission spectra. Other approaches used the spectra of Fabry-Perot resonances in cavities [37-39], time of flight measurements using an external Kerr shutter [40], or direct measurements employing NSOM [14]. UMPS gives access to the spectrally resolved change of transmission (ΔT/T) as displayed in Fig. 3b for a position 1.8 mm from the beginning of the investigated waveguide. For positive delay times, dark colours indicate a decrease in transmission due to absorption. Probe light arriving before the pump excitation, i.e. at negative delay times, is not modulated, resulting in a sharp step edge in the modulation time response. The pump-probe map shows two step edges, corresponding to TE and TM polarized light. The latter is weakly confined to the waveguide, and contributes significantly in our experiment only in a limited spectral range around the centre wavelength of the probe. Due to group index dispersion, the arrival time of the probe light is a function of wavelength and the step edges are not vertical. By following the absolute position and the evolution of the slope of the step edge over the length of the waveguide (Fig. 3c,d) the group index and its linear dispersion can be calculated. Since only the evolution of the slope and not the absolute slope of the step edge is analysed along the waveguide, this technique is not sensitive to chirp in the incident probe pulses. The resulting group index dispersions are shown in Fig. 3e and f for the TE and TM polarization components respectively (blue lines). Good agreement is obtained with numerical results based on finite element method (FEM) simulations (red lines), with the small systematic shift in TM most likely representing a difference in precise design geometry. Our measurements are also in agreement with the work of Dulkeith et al., who investigated comparable waveguides [41].

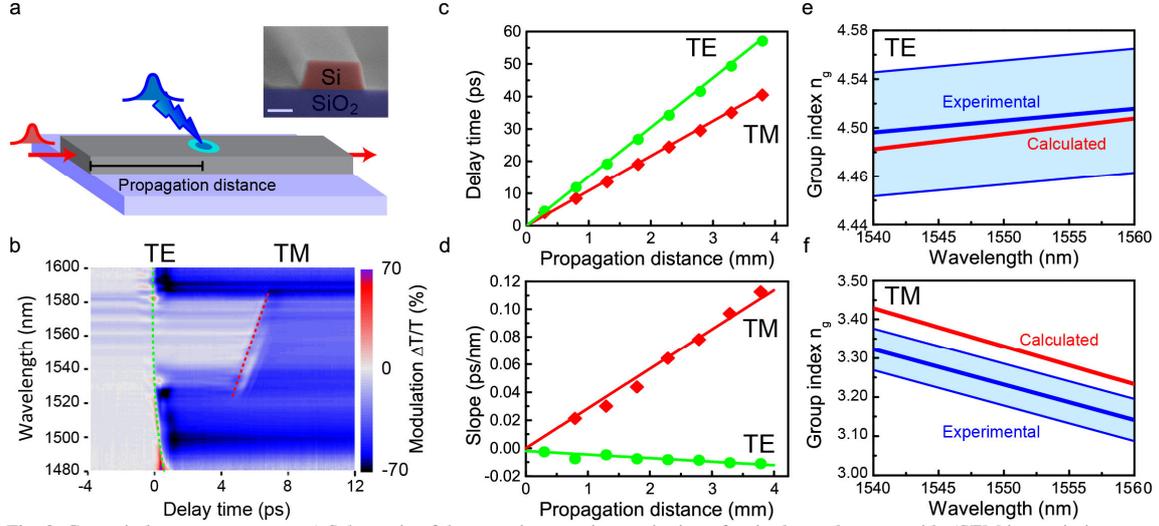

Fig. 3. Group index measurements. a) Schematic of the experiment using excitation of a single-mode waveguide (SEM image in inset, scale bar 200 nm). b) Delay time scan for a straight waveguide, performed 1.8 mm from waveguide beginning (Zero delay time in this scan chosen for clarity as the arrival time of the TE light). Red and green dashed lines correspond to arrival times of light propagating in the TE and TM modes, respectively. c,d) Absolute arrival time c) and slope d) of the TE and TM components extracted from UPMS experiments at different propagation distances in the waveguide, for a wavelength of 1550 nm. e,f) Linear group index dispersion obtained from the slopes of c,d) for TE e) and TM f) modes. Red lines indicate calculated group index using a numerical model (see Methods).

**Time response of a ring resonator.** Typically, quality factors of ring resonators are obtained from the linear transmission spectrum using the equation $Q = f/\Delta f$ on the dips in transmission (e.g. [42-44]). However, closer to the fundamental definition of the Q-factor as the ratio of energy stored to energy dissipated per cycle, are measurements in the time-domain. On resonance, light from the bus waveguide is coupled into the ring with high efficiency and is stored over many cycles before leaking back to the output waveguide. The corresponding delay time can be detected by UPMS using a negative delay time, i.e. probe pulses arriving at a time before the pump pulses. In this case, only light stored in the ring resonator for at least the delay time is affected by the excitation and the corresponding photomodulation response will be a measure of the amount of energy stored in the ring. Eventually, an exponentially decaying UPMS effect is expected, as less and less light from the ring transfers back to the bus waveguide.

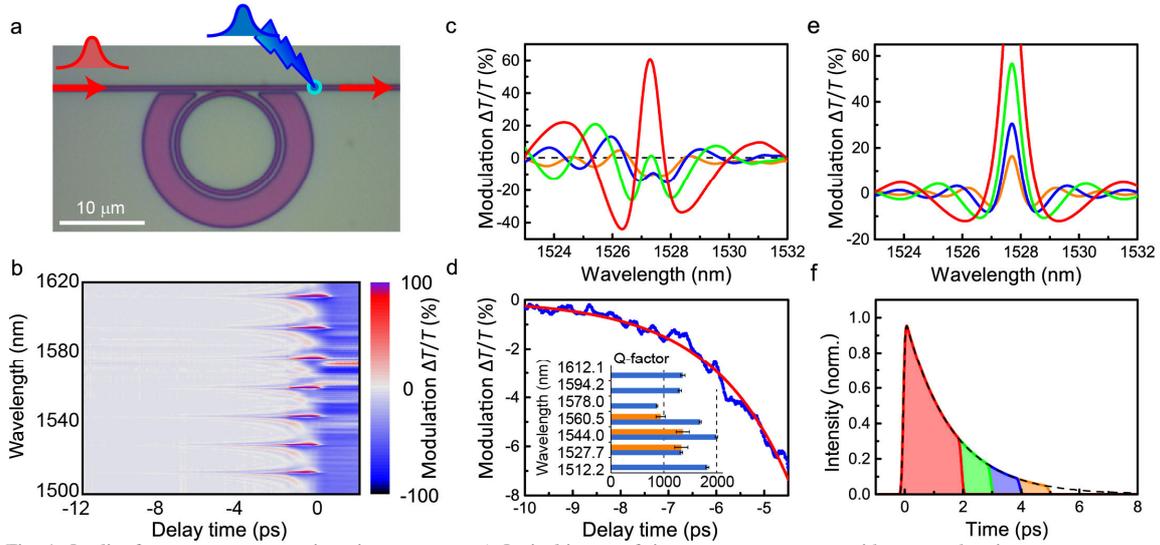

Fig. 4. Quality factor measurements in a ring resonator. a) Optical image of ring resonator structure with arrows showing arrangement for light incoupling and excitation. b) Delay time scan of change in transmission (ΔT/T) of probe pulses on the output bus waveguide, shortly after the ring resonator. The scan shows the spectral features of the ring resonator at negative delay times, due to energy storage and ringdown. c) Slices through b) around λ = 1523-1532 nm and for delay times of -5 ps (orange), -4 ps (blue), -3 ps (green), and -2 ps (red). d) Modulation at resonance of c) against delay time. Red line indicates single-exponential decay fit with 1.63 ps photon dwell time in the ring resonator. Inset: histogram of quality factors obtained from UMPS experiments presented in b) (blue) and compared with quality factors from transmission measurements (orange). e,f) Calculated modulation spectra (e) assuming exponential ringdown (f, dashed black line) with ultrafast photomodulation gate included after 2 ps (red), 3 ps (green), 4 ps (blue) and 5 ps (orange).

For UPMS characterization of the quality factor, we selected a pump spot on the bus waveguide after the ring resonator as shown in Fig. 4a. Figure 4b shows the spectrally resolved change of transmission (Δ$T$/$T$) for this investigated position as function of delay time. The resonances of the ring are clearly visible in the graph at negative delay times. Similar to the straight waveguide, the zero timing slightly depends on the wavelength due to group index dispersion in the bus waveguide. Figure 4c represents cuts through the graph in Fig. 4b at the resonance wavelength of 1527.7 nm for delay times between -5 ps and -2 ps. We find an oscillating modulation which is symmetric around the resonance wavelength. As we are pumping only the bus waveguide, no shift in the cavity resonance is observed, and the symmetric sidebands can be attributed to an ultrafast amplitude modulation of the transmitted component due to photoabsorption, which will be explained further below. In order to obtain the cavity ringdown time, we plotted in Fig. 4d the modulation amplitude at the resonance wavelength versus delay time. A single-exponential decay was observed for long negative delay times beyond -4 ps. The different resonances of the ring within the spectral range of our setup were analysed and corresponding Q factors are presented in a histogram in the inset of Fig. 4d (blue bars). Values of Q-factors in the range 1000 to 2000 are observed, which is of the same order as Q-factors obtained from optical transmission spectra of the same device using a narrowband telecom setup (orange bars). Quality factor measurements obtained from transmission data are generally susceptible to noise, which in particular for the end-face coupling used in our experiments affects the accuracy of this method. In comparison, time-domain UPMS offers a more direct way of determining the Q-factor through the characteristic ringdown time of the resonance. Further, effects stemming from reflections on interfaces or any defects/deviations which might be present in the device and contaminate the transmission spectrum are excluded in differential $\Delta T/T$ measurements.

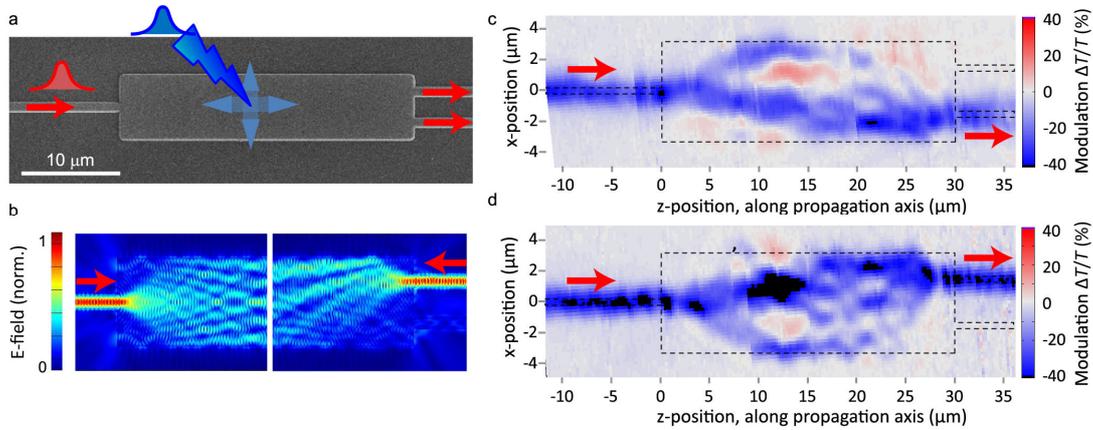

Fig. 5. Mode pattern mapping in a multimode interference (MMI) device. a) micrograph of the studied system. b) FDTD simulation intensity maps. In the left part, the incident wave is launched from the input waveguide. Areas of low intensity correspond to measurements presented in c,d. In the right part, the wave is incident from one output waveguide. The light paths contributing to this output are visible. c,d) UMPS measurements: spatial photomodulation maps of change in transmission ($\Delta T/T$) of probe pulses through the two outputs respectively, if the MMI is illuminated with the pump pulses at the position given by the map ($\lambda = 1556.5$ nm). The pump pulses arrive 4 ps before the probe pulses.

The oscillating modulation spectrum in Fig. 4c can be understood from amplitude modulation caused by the ultrafast photomodulation gate. Figure 4e reproduces the general trend, namely an oscillating sideband with a characteristic frequency bandwidth, which is reduced towards longer negative delay times of the gate. The modulation spectra were obtained by Fourier transforming the time-response functions shown in Fig. 4f. The unperturbed ring resonator response is modelled as a carrier frequency with exponentially decaying envelope as indicated by the black dashed line in Fig. 4f, resulting in a single Lorentzian resonance at 1527.7 nm. UPMS is subsequently included by suppressing the transmission after a time corresponding to the delay time of Fig. 4b. This truncation produces the characteristic ripple in the frequency spectrum, as is shown in the differential $\Delta T/T$ response in Fig. 4e and which is found experimentally.

**Mapping of a multimode interference (MMI) splitter.** We use UPMS to investigate the spatial distribution of light involved in the transport from the input to one of the output waveguides in an MMI. In MMI-devices, a complex mode pattern defines how light is coupled to the output waveguides. By introducing a localized modulation in the MMI-region, the contribution of an exited spot to the selected output can be altered, and thus identified. Generally, two effects will be induced by the laser excitation: (i) rerouting of light from one output to the other and (ii) localized suppression of transmission due to absorption or scattering. Figure 5a shows an SEM image of the MMI device under study. As shown in the left part of Fig. 5b, the field distribution in the symmetric 50/50 splitter device calculated by finite-difference time domain (FDTD) simulations shows a symmetric profile. However, the distribution contributing to only one of the two outputs is asymmetric, as can be appreciated by looking at the time-reversed situation, i.e. light launched from one of the two inputs (right part of Fig. 5b). In contrast to NSOM, which provides the symmetric mode intensity, UPMS selectively provides specific information on the light paths involved in the light transport to the detection point. Figure 5d,e show photomodulation maps corresponding to the transmission measured at the lower and upper output waveguides, respectively, at a wavelength of 1556.5 nm. Blue regions correspond to the case of reduced transmission. Comparison with the time-reversed asymmetric profile of Fig. 5b shows strong similarities between the regions of suppressed transmission and high intensity paths. Next to regions of reduced transmission, regions are observed where the transmission is enhanced. These red areas correspond to parts of the mode profile where the contribution to the selected output can be increased. For such regions, the induced change in $\Delta n_{Si}$ redirects light from one waveguide to the other waveguide. As the average magnitude of blue areas is significantly larger than for red areas, a major contribution to the reduction in transmission for blue areas can be attributed to induced absorption and scattering of light out of individual lightpaths. For rerouting of light, similar magnitudes for red and blue areas would be expected.

**Conclusions**

We have developed a new technique of ultrafast photo-modulation spectroscopy (UPMS). This technique offers a multidimensional, hyperspectral characterization for silicon waveguide based elements at the device level. By introducing a localized modulation in the refractive index of silicon employing a short pump wavelength of 400 nm, light propagation in silicon photonic elements can be altered and identified with a spatial resolution that compares with telecommunication wavelengths inside the high-index structures. A decrease in the refractive index of silicon due to optical pumping larger than 0.4 in the real part and an increase of up to 0.06 in the imaginary part is obtained at medium pump fluences of around 100 pJ/µm$^2$, without inducing any damage to the silicon nanophotonic elements. This large modulation of the real part of the refractive index on a time scale of picoseconds is of interest for active control of nanophotonic devices, which will be explored in future work. UPMS provides access to time-domain measurements and allows for simultaneous spatial and spectral mapping. Compared to scanning near-field techniques, the far-field excitation is robust and can be used in the presence of transparent cladding layers. We have demonstrated the versatility for imaging and quantitative analysis of essential components such as MZIs, waveguides, ring resonators, and MMI devices. While we have focused here on the characterization aspects, the high modulation effects UPMS can also be used to efficiently modulate silicon nanophotonic devices on length scales of only a few optical wavelengths, enabling new avenues toward reconfigurable integrated circuits with small footprints.

**Methods**

**Optical setup.** The laser system used for the presented measurements is a Ti-Sapphire laser coupled with an optical parametric amplifier giving 150 fs pulses at a 250 kHz repetition rate. It provides an output for 400 nm pump pulses and an output of variable wavelength (visible range to 2 µm) for probe pulses. All presented measurements used a central wavelength of 1550 nm for the probe pulses and TE-polarization. Pump pulse energies for the application examples varied between 0.08 nJ and 0.24 nJ. In our setup (see Fig. 1) the probe pulses are coupled by a microscope objective into an input waveguide. The transmitted light of one output of the DUT is coupled into a fibre and subsequently to a spectrometer. The pump pulses can be focused from the top onto any position of the DUT.

**Mach-Zehnder Interferometer.** The asymmetric MZI is composed from rib waveguides (450 nm width, 220 nm total thickness, 100 nm slab thickness, silicon dioxide as lower and top cladding). The short arm of the MZI has a length of around 1040 µm, the long arm is 180 µm longer. While this path length difference is longer than the length of the femtosecond pulses, it is much less than the millimeter coherence length associated with the spectrometer resolution, therefore clear interference fringes were obtained in the transmission spectrum. Changes in the real part of the effective index $\text{Re}(\Delta n_{eff})$ were obtained from the fringe shift $\Delta\lambda$ through the relation $\text{Re}(\Delta n_{eff})L_p = \frac{\lambda}{\lambda_{FSR}}\Delta\lambda$, with the free spectral range given by $\lambda_{FSR} = \frac{\lambda^2}{n_{eff}\Delta L}$. Here, $n_{eff}$ denotes the effective mode index, $\lambda$ the wavelength in vacuum and $\Delta L$ the length difference between the arms of the MZI. The pumping length $L_p$ corresponds to the length of the pumped section of the waveguide, where we assumed a homogeneous pumping intensity. The change in the imaginary part of $\Delta n_{eff}$ was obtained from the amplitude of the interference fringes with ($A'$) and without ($A$) pumping through the relation $\text{Im}(\Delta n_{eff})L_p = -\frac{\lambda}{2\pi}\ln\frac{A'}{A}$. The measured changes in the effective index of the waveguide $\Delta n_{eff}$ were translated to changes in the refractive index of silicon $\Delta n_{Si}$ by modelling the waveguide using the finite-element method (FEM). The model assumed that the surrounding silicon dioxide cladding does not change. Two measurements with different pump pulse spot diameters were performed, corresponding to $L_p$=1.5 µm and 3.5 µm. The pumping lengths were adjusted by defocusing the microscope objective and were measured by lateral scans over the waveguide. We calculated the effect lengths as the full width at 1/e maximum of a Gaussian fit to the lateral scans minus the width of the waveguide. The thick solid lines in Fig. 2c denote a linear fit for the imaginary part and a fit of $\text{Re}(\Delta n_{eff}) = c\frac{p}{p+S}$ for the real part, with scaling factor $c$ and saturation parameter $S$=56±12 pJ/µm².

**Waveguide.** The group index measurements were performed on a straight wire waveguide (470 nm width, 220 nm thickness, silicon dioxide lower cladding and air as top cladding) of about 4 mm length. The error bars in Fig. 3e,f are 95% confidence levels for the linear fits on the measured step edge positions and slopes. For the numerical calculation of the group index, the waveguide cross section of the waveguide (see inset in Fig. 3a) was measured using SEM and implemented in the FEM model.

**Ring resonator.** The ring resonator and the bus waveguide are made from identical waveguides as the MZI. The decay time of the ring resonator was obtained using exponential fitting of the (negative) modulation amplitude at resonance for PPTDs between -10 ps and -4 ps (in this range single-exponential decay was observed). The time constant τ of this exponential decay was converted to the Q-factor using the relation [45]:

$$Q = \frac{2\pi c\tau}{\lambda} \qquad (1)$$

**Multimode interference device.** Spatial maps of the MMI were obtained for both output waveguides by translating the output fibre to pick up light from the individual outputs. In this way, the symmetry of the device was proven to good agreement. Maps of the MMI for various input conditions were calculated using the Lumerical FDTD package.

# References

[1] Soref, R. Silicon Photonics: A Review of Recent Literature. *Silicon* **2**, 1–6 (2010).

[2] Passaro, V.M.N. et al. Recent Advances in Integrated Photonic Sensors. *Sensors* **12**, 15558-15598 (2012).

[3] Jalali, B. & Fathpour, S. Silicon Photonics. *J. of Lightw. Techn.* **24**, 4600-4614 (2006).

[4] Yamada, K. at al. High-performance silicon photonics technology for telecommunications applications. *Sci. Technol. Adv. Mater.* **15**, 024603 (2014).

[5] Beaud, P. et al. Optical Reflectometry with Micrometer Resolution for the Investigation of Integrated Optical Devices. *J. Quant. El.* **25**, 755-759 (1989).

[6] Youngquist, R.C., Carr, S., & Davies D.E.N. Optical coherence-domain reflectometry: a new optical evaluation technique. *Opt. Let.* **12**, 158-160 (1987).

[7] Kohlhaas, A., Frömchen, C. & Brinkmeyer, E. High-Resolution OCDR for Testing Integrated-Optical Waveguides: Dispersion-Corrupted Experimental Data Corrected by a Numerical Algorithm. *J. Lightw. Techn.* **9**, 1493-1502 (1991).

[8] Glombitza, U. & Brinkmeyer, E. Coherent Frequency-Domain Reflectometry for Characterization of Single-Mode Integrated-Optical Waveguides. *J. Lightw. Techn.* **11**, 1377-1384 (1993).

[9] Loncar, M. et al. Experimental and theoretical confirmation of Bloch-mode light propagation in planar photonic crystal waveguides. *Appl. Phy. Lett.* **80**, 1689-1691 (2002).

[10] McNab, S.J., Moll, N. & Vlasov, Y.A. Ultra-low loss photonic integrated circuit with membrane-type photonic crystal waveguides. *Opt. Ex.* **22**, 2927-2939 (2003).

[11] Hopman, W.C.L., Hoekstra, H.J.W.M., Dekker, D., Zhuang, L. & de Ridder, R.M. Far-field scattering microscopy applied to analysis of slow light, power enhancement, and delay times in uniform Bragg waveguide gratings. *Opt. Ex.* **15**, 1851-1870 (2007).

[12] García, P.D., Smolka, S., Stobbe, S. & Lodahl, P. Density of states controls Anderson localization in disordered photonic crystal waveguides. *Physical Review B* **82**, 165103 (2010).

[13] Sapienza, R. et al. Deep-subwavelength imaging of the modal dispersion of light. *Nature Materials* **11**, 781–787 (2012).

[14] Gersen, H., et al. Real-space observation of ultraslow light in photonic crystal waveguides. *Phys. rev. let.* **94**, 073903 (2005).

[15] Engelen, R.J.P. et al. Ultrafast evolution of photonic eigenstates in k-space. *Nature Physics* **3**, 401-405 (2007).

[16] Kramper, P. Near-field visualization of light confinement in a photonic crystal microresonator. *Opt. Lett.* **29**, 174-176 (2004).

[17] Koenderink, A.F., Kafesaki, M., Buchler, B.C. & Sandoghdar, V. Controlling the Resonance of a Photonic Crystal Microcavity by a Near-Field Probe. *Phys. Rev. Lett.* **95**, 153904 (2005).

[18] Märki, I., Salt, M., Herzig, H. P. Tuning the resonance of a photonic crystal microcavity with an AFM probe. *Opt. Exp.* **14**, 2969-2978 (2006).

[19] Hopman, W.C.L. et al. Nano-mechanical tuning and imaging of a photonic crystal micro-cavity resonance. *Opt. Exp.* **14**, 8745-8752 (2006).

[20] Robinson, J.T., Preble, S.F., Lipson, M. Imaging highly confined modes in sub-micron scale silicon waveguides using Transmission-based Near-field Scanning Optical Microscopy. *Opt. Exp.* **14**, 10588-10595 (2006).

[21] Burresi, M. et al. Probing the Magnetic Field of Light at Optical Frequencies. *Science* **326**, 550-553 (2009).

[22] Gersen,H., Korterik, J.P., van Hulst, N.F. & Kuipers., L. Tracking ultrashort pulses through dispersive media: Experiment and theory. *Phys. Rev. E* **68**, 026604 (2003).

[23] Beggs, D.M. et al. Ultrafast Tunable Optical Delay Line Based on Indirect Photonic Transitions. *Phys. Rev. Lett.* **108**, 213901 (2012).

[24] Beggs, D.M., Krauss, T.F., Kuipers, L., & Kampfrath, T. Ultrafast Tilting of the Dispersion of a Photonic Crystal and Adiabatic Spectral Compression of Light Pulses. *Phys. Rev. Lett.* **108**, 033902 (2012).

[25] Opheij, A. et al. Ultracompact (3 μm) silicon slow-light optical modulator. *Scientific Reports* **3**, 3546 (2013).

[26] Liang, T.K. et al. Ultrafast nonlinear optical studies of silicon nanowaveguides. *Opt.Exp.* **13**, 7298-7303 (2005).

[27] Slavik, R., Park, Y., Kulishov, M., Morandotti, R. & Azana, J. Ultrafast all-optical differentiators. *Opt.Exp.* **14**, 10699-10707 (2006).

[28] Preble, S.F., Xu, Q., Schmidt, B.S. & Lipson M. Ultrafast all-optical modulation on a silicon chip. *Opt. Lett.* **30**, 2891-2893 (2005).

[29] Upham, J. et al The capture, hold and forward release of an optical pulse from a dynamic photonic crystal nanocavity. *Opt. Exp.* **21**, 3809-3817 (2013).

[30] Kondo, K. et al. Ultrafast Slow-Light Tuning Beyond the Carrier Lifetime Using Photonic Crystal Waveguides, *Phys. Rev. Lett.* **110**, 053902 (2013).

[31] Motamedi, A.R., Nejadmalayeri, A.H., Khilo, A., Kärtner, F.X. & Ippen, E. P. Ultrafast nonlinear optical studies of silicon nanowaveguides. *Opt. Exp.* **20**, 4085-4101 (2012).

[32] Soref, R.A. & Bennett, B. R. Electrooptical effects in silicon. *IEEE J. Quant. Electr.* **23**, 123-129 (1987).

[33] Sokolowski-Tinten, K., von der Linde, D., Generation of dense electron-hole plasmas in silicon. *Phys. Rev. B* **61**, 2643-2650 (2000).

[34] Sabbah, A. J., & Riffe, D. M. (2002). Femtosecond pump-probe reflectivity study of silicon carrier dynamics. *Phys. Rev. B* **66**, 165217 (2002).

[35] Kampfrath, T. Ultrafast rerouting of light via slow modes in a nanophotonic directional coupler. *Appl. Phys. Lett.* **94**, 241119 (2009).

[36] Vlasov, Y.A., O'Boyle, M., Hamann, H.F. & McNab, S.J. Active control of slow light on a chip with photonic crystal waveguides. *Nature* **483**, 65-69 (2005).

[37] Letartre, X. et al. Group velocity and propagation losses measurement in a single-line photonic-crystal waveguide on InP membranes. *Appl. Phys. Let.* **79**, 2312-2314 (2001).

[38] Lee, M.W. et al. Characterizing photonic crystal waveguides with an expanded k-space evanescent coupling technique. *Opt. Ex.* **16**, 13800-13808 (2008).

[39] Notomi, M. et al. Extremely Large Group-Velocity Dispersion of Line-Defect Waveguides in Photonic Crystal Slabs. *Phys. Rev. Let.* **87**, 253902 (2001).

[40] Finlayson, C.E. et al. Slow light and chromatic temporal dispersion in photonic crystal waveguides using femtosecond time of flight. *Phys. Rev. E* **73**, 016619 (2006).

[41] Dulkeith, E., Xia, F., Schares, L., Green, W.M.J. & Vlasov, Y.A. Group index and group velocity dispersion in silicon-on-insulator photonic wires. *Opt. Ex.* **14**, 3853-3863 (2006).

[42] Niehusmann, J. et al. Ultrahigh-quality-factor silicon-on-insulator microring resonator. *Opt. Let.* **29**, 2861-2863 (2004).

[43] Baehr-Jones, T., Hochberg, M., Walker, C. & Scherer, A. High-Q optical resonators in silicon-on-insulator-based slot waveguides. *Appl. Phys. Let.* **86**, 081101 (2005).

[44] Almeida, V.R., Barrios, C.A., Panepucci, R.R. & Lipson, M. All-optical control of light on a silicon chip. *Nature* **431**, 1081-1084 (2004).

[45] Xu, Q. & Lipson, M. All-optical logic based on silicon micro-ring resonators. *Opt. Ex.* **15**, 924-929 (2007).

**Acknowledgements**

RB and OLM acknowledge support from EPSRC through grant no. EP/J016918. The authors thank G. Mashanovich of the Optoelectronics Research Centre, University of Southampton, for stimulating discussions and for proofreading of the manuscript.